\RequirePackage{fix-cm}
\documentclass[smallextended]{svjour3}       
\smartqed  
\usepackage{graphicx}
\begin{document}

\title{Lightning black holes as unidentified TeV sources
}


\author{Kouichi Hirotani, 
        Hung-Yi Pu \and Satoki Matsushita
}


\institute{Kouichi Hirotani and Satoki Matsushita \at
           11F of AS/NTU Astronomy-Mathematics Building, 
           No.1, Sec. 4, Roosevelt Rd, Taipei 10617, Taiwan, R.O.C.  \\
              Tel.: +886-2-2366-5406\\
              Fax:  +886-2-2367-7849\\
              \email{hirotani@asiaa.sinica.edu.tw}           
           \and
           Hung-Yi Pu \at
           Perimeter Institute for Theoretical Physics,
           31 Caroline Street North, Waterloo, ON, N2L, 2Y5, Canada
}

\date{Received: date / Accepted: date}

\maketitle

\begin{abstract}
Imaging Atmospheric Cherenkov Telescopes have revealed more than 
100 TeV sources along the Galactic Plane, around 45\% of them 
remain unidentified. 
However, radio observations revealed that dense molecular clumps 
are associated with 67\% of 18 unidentified TeV sources. 
In this paper, we propose that an electron-positron
magnetospheric accelerator emits detectable TeV gamma-rays
when a rapidly rotating black hole enters a gaseous cloud. 
Since the general-relativistic effect plays an essential role 
in this magnetospheric lepton accelerator scenario, 
the emissions take place in the direct vicinity of the event horizon, 
resulting in a point-like gamma-ray image.
We demonstrate 
that their gamma-ray spectra have two peaks around 0.1 GeV and 0.1 TeV 
and that the accelerators become most luminous when the mass accretion 
rate becomes about 0.01\% of the Eddington accretion rate.
We compare the results with alternative scenarios
such as the cosmic-ray hadron scenario,
which predicts an extended morphology of the gamma-ray image
with a single power-law photon spectrum from GeV to 100~TeV.
%
\end{abstract}

\section{Introduction}
\label{intro}
The Imaging Atmospheric Cherenkov Telescopes (IACTs) provides 
a wealth of new data on various energetic astrophysical objects, 
increasing the number of detected very-high-energy (VHE) gamma-ray sources,
typically between 0.01 and 100 TeV, 
from 7 to more than 200 in this century
\footnote{TeV Catalog (http:www.tevcat.uchicado.edu)}. 
Among the presently operating three IACTs, 
High Energy Stereoscopic System (HESS) \cite{wilhelmi09}
has so far discovered 42 new 
VHE sources along the Galactic Plane, 22 of which are still unidentified. 
The nature of these unidentified VHE sources may be hadronic origin
\cite{black73,issa81}, 
because protons can be efficiently accelerated into VHE in a supernova 
remnant to penetrate into adjacent dense molecular clouds, which leads to 
an extended gamma-ray image. By a systematic comparison between 
the published HESS data and the molecular radio line data, 
38 sources are found to be associated with dense molecular 
clumps out of the 49 Galactic VHE sources covered 
by 12 mm observations \cite{dewilt17}. 

There is, however, an alternative scenario for the VHE emissions 
from gaseous clouds. 
In the Milky Way, molecular gas is mostly located in giant molecular clouds, 
in which massive stars are occasionally formed. 
If a massive star evolves into a black hole and encounters 
an adjacent molecular clouds, it accretes gases. 
It is, therefore, noteworthy that a rapidly rotating, 
stellar-mass black hole emits copious gamma-rays in 0.001-1 TeV
\cite{hiro16a}, 
provided that its dimensionless accretion rate 
$\dot{m} \equiv \dot{M}/\dot{M}_{\rm Edd}$, 
satisfies $6 \times 10^{-5} < \dot{m} < 2 \times 10^{-4}$, 
where $\dot{M}$ designates the mass accretion rate, 
$\dot{M}_{\rm Edd} \equiv 1.39 \times 10^{19} M_1 \mbox{g s}^{-1}$ 
is the Eddington accretion rate,
$M_1= M/(10 M_\odot)$ and $M_\odot$ denotes the solar mass. 
The electric currents flowing in such an accreting plasma create 
the magnetic field threading the event horizon. 
In this leptonic scenario, 
migratory electrons and positrons ($e^{\pm}$’s) are accelerated to TeV 
by a strong electric field exerted along these magnetic field lines, 
and cascade into many pairs as a result of 
the collisions between 
the VHE photons emitted by the gap-accelerated $e^\pm$'s and
the IR photons emitted by the hot $e^-$'s in the equatorial accretion flow.
The resulting gamma radiation takes place only near the black hole; 
thus, their VHE image should have a point-like morphology 
with a spectral turnover around TeV. 

\section{Black hole accretion in a gaseous cloud}
\label{sec:1}
\begin{figure}
 \includegraphics[width=11.8 cm]{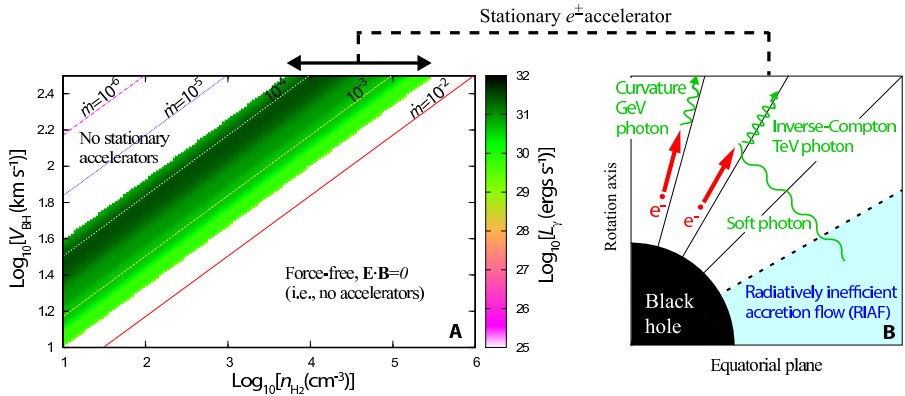}
\caption{
(A) Luminosity of black-hole lepton accelerators when a ten-solar-mass, 
extremely rotating ($a=0.99 r_{\rm g}$) black hole is moving 
with velocity $V$ in a cloud with molecule hydrogen density $n_{\rm H_2}$. 
For an atomic hydrogen gas with density $n_{\rm HI}$, 
put $n_{\rm H_2}=n_{\rm HI}/2$, because the mass is halved. 
The five straight lines correspond to the Bondi-Hoyle accretion rates, 
$10^{-2}$, $10^{-3}$, $10^{-4}$, $10^{-5}$, and $10^{-6}$, as labeled. 
In the lower-right white region, 
the enhanced photon illumination from the equatorial accretion flow
results in an efficient pair production, and hence
a complete screening of the magnetic-field-aligned electric field;
thus, the accelerator vanishes in this region.
In the upper-left white region, stationary accelerators cannot be 
formed (see text). 
Thus, stationary accelerators arise only in the green-black region. 
(B)
Schematic figure (side view) of a black-hole magnetosphere. 
The polar funnel is assumed to be bounded from the radiatively 
inefficient accretion flow (cyan region) at colatitude $\theta=60^\circ$
(dashed line) from the rotation axis (ordinate).
}
\label{fig:1}       
\end{figure}

When a black hole moves in a gaseous cloud, the particles are captured 
by the hole's gravity to form an accretion flow. 
Since the temperature is very low in a molecular cloud, the black hole 
will move with a supersonic velocity $V$, forming a bow shock behind. 
Under this situation, the gas pressure can be neglected and the particles 
within the impact parameter $r_{\rm B} \sim GM V^{-2}$ 
from the black hole will be captured. 
For a homogeneous gas, the mass accretion rate becomes \cite{bondi44}
$\dot{M}_{\rm B} = 4\pi\lambda(GM)^2 (C_{\rm S}{}^2+V^2)^{-3/2} \rho
 \approx 4\pi\lambda(GM)^2 V^{-3} \rho $, 
where $\rho$ denotes the mass density of the gas,  
$\lambda$ a constant of order unity, 
$G$ the gravitational constant,
and $C_{\rm S}$ the sound speed in the homogeneous gas;
the last near equality comes from the supersonic nature
(i.e., $V \gg C_{\rm S}$) of accretion.
For a molecular hydrogen gas, 
we obtain the dimensionless Bondi accretion rate
$\dot{m}_{\rm B} \equiv \dot{M}_{\rm B}/\dot{M}_{\rm Edd}
 = 5.4 \times 10^{-9} \lambda n_{\rm H_2} M_1 (V/10^2 \mbox{km s}^{-1})^{-3}$, 
where $n_{\rm H_2}$ denotes the number density of hydrogen molecules per 
$\mbox{cm}^3$. 
Representative values of $\dot{m}_{\rm B}$ are plotted 
as the five straight lines in figure~\ref{fig:1}A. 

Since the accreting gases have little angular momentum as a whole 
with respect to the black hole, they form an accretion disk 
only within a radius that is much less than $r_{\rm B}$. 
Thus, we neglect the mass loss as a disk wind between $r_{\rm B}$ and 
the inner-most region, and evaluate the accretion rate 
near the black hole, $\dot{m}$, with $\dot{m}_{\rm B}$. 
In what follows, we consider a ten solar mass black hole, 
which is typical as a stellar-mass black hole 
\cite{tararen16,corral16}. 
It is reasonable to suppose that such black holes have 
kick velocities of $V<10^2 \mbox{km s}^{-1}$ 
with respect to the star-forming region.
Under this circumstance, a typical velocity dispersion 
in a molecular cloud, $\Delta V < 10 \mbox{ km s}^{-1}$,
is an order of magnitude less than $V$.
Accordingly, the net specific angular momentum of the gas at $r_{\rm B}$,
which is typically $r_{\rm B} \Delta V$,
is much less than the Keplerian value, 
$r_{\rm B} V = \sqrt{GM r_{\rm B}}$.
At a much smaller radius $r=(\Delta V/V)^2 r_{\rm B} < 0.01 r_{\rm B}$,
$r_{\rm B} \Delta V$ equals the Keplerian value;
therefore, a disk is formed within this radius.
Since the accreting gas does not have to lose angular momentum
when falling from $r=r_{\rm B}$ to $(\Delta V/V)^2 r_{\rm B}$,
we neglect the mass loss in this region
and evaluate the accretion rate $\dot{m}$ near the BH
with the Bondi-Hoyle accretion rate $\dot{m}_{\rm B}$ for simplicity
in the present paper.

\section{Development of charge-starved magnetosphere}
\label{sec:2}
When $\dot{m}$ becomes typically less than $10^{-2}$, 
Coulomb collisions become so inefficient that the accreting protons’ 
thermal energy cannot be efficiently transferred to the electrons. 
If the accretion rate decreases to $\dot{m}<10^{-2.5}$, 
such a radiatively inefficient accretion flow (RIAF) 
\cite{ichimaru79,narayan94,mahad97}
cannot supply enough soft gamma-rays that are needed to sustain 
the magnetosphere force-free \cite{levinson11}. 
Accordingly, a charge-starved, nearly vacuum magnetosphere develops 
in the polar funnel (fig.~\ref{fig:1}B), 
because the equatorial accreting plasmas cannot 
penetrate there due to the centrifugal-force barrier. 
If the accretion rate further decreases to $\dot{m}<6 \times 10^{-5}$, 
stationary pair production cascade cannot be sustained. 
However, the luminosity of such a non-stationary accelerator 
becomes less than the stationary cases, 
because only a weaker magnetic field can be confined near 
the black hole for a lower accretion rate. 
Thus, we consider only the range  
$6 \times 10^{-5} < \dot{m} < 2 \times 10^{-4}$
(green-black region in fig.~\ref{fig:1}A) 
and concentrate on stationary accelerators. 
It is noteworthy that if a stellar-mass black hole moves slowly 
(i.e., $V \ll 10^2 \mbox{km s}^{-1}$), 
its accelerator can be activated with a low gas density 
(e.g., $n_{\rm H_2} \ll 10^3 \mbox{cm}^{-3}$; fig.~\ref{fig:1}A). 
Thus, a significant gamma-ray emission is possible 
when a black hole encounters not only a dense molecular cloud 
but also a diffuse molecular gas or even an atomic gas. 

\section{Lepton accelerator in black hole magnetospheres}
\label{sec:3}
In a vacuum magnetosphere, an electric field, $E_\parallel$, 
arises along the magnetic field lines. 
Accordingly, electrons and positrons (red arrows in fig.~\ref{fig:1}B) 
are accelerated into ultra-relativistic energies to emit 
high-energy gamma-rays (wavy line with middle wavelength) 
via the curvature process 
(a kind of the synchrotron process whose the electron's
 gyro radius is replaced with the macroscopic curvature radius
 of three-dimensional electron's motion)
and VHE gamma-rays (wavy line 
with shortest wavelength) via the inverse-Compton (IC) scatterings 
of the soft photons (wavy line with longest wavelength) 
emitted from the RIAF. 
A fraction of such VHE photons collide with the soft RIAF photons 
to materialize as $e^\pm$ pairs, which partially screen the original 
$E_\parallel$ when they separate. 
It is noteworthy that pair annihilation is negligible compared 
to pair production in BH gaps.
To compute the actual strength of $E_\parallel$, 
we solve the $e^\pm$ pair production cascade 
in a stationary and axisymmetric magnetosphere 
on the meridional plane ($r$,$\theta$), 
where $r$ denotes the Boyer-Lindquist radial coordinate, 
and $\theta$ does the colatitude measured from the rotation axis.
The black hole’s rotational energy is electromagnetically extracted 
via the Blandford-Znajek process \cite{BZ77}
and partially dissipated 
as particle acceleration and the resultant radiation within the accelerator. 
It is noteworthy that the electrodynamics of this lepton accelerator 
is essentially described by the general-relativistic Goldreich-Julian 
charge density, which is governed by the magnetic-field 
strength and the frame-dragging effects. 
Thus, the accelerator solution little depends on the magnetic field 
configuration near the event horizon. 
We therefore assume that the magnetic field is radial 
in the meridional plane and that magnetic axis is aligned 
with the rotation axis. 
The magnetic field lines are twisted in the azimuthal direction 
due to the frame-dragging effect, and its curvature radius is 
assumed to be $r_{\rm g}$ in the local reference frame. 
This assumption modestly affects the curvature spectrum, 
but does not affect the entire electrodynamics, 
because the pair-production process, 
and hence the screening of $E_\parallel$ is governed by the highest-energy, 
IC-scattered photons. 

\section{Basic equations}
\label{sec:4}
Let us quantify the accelerator electrodynamics.
In a rotating black-hole magnetosphere,
electron-positron accelerator is formed in the direct vicinity
of the event horizon.
Thus, we start with describing the background spacetime
in a fully general-relativistic way.
We adopt the geometrized unit, putting $c=G=1$, where
$c$ and $G$ denote the speed of light and the gravitational constant,
respectively.
Around a rotating BH, 
the spacetime geometry is described by the Kerr metric
\cite{kerr63}.
In the Boyer-Lindquist coordinates, it becomes 
\cite{boyer67} 
\begin{equation}
 ds^2= g_{tt} dt^2
      +2g_{t\varphi} dt d\varphi
      +g_{\varphi\varphi} d\varphi^2
      +g_{rr} dr^2
      +g_{\theta\theta} d\theta^2,
  \label{eq:metric}
\end{equation}
where 
\begin{equation}
   g_{tt} 
   \equiv 
   -\frac{\Delta-a^2\sin^2\theta}{\Sigma},
   \qquad
   g_{t\varphi}
   \equiv 
   -\frac{2Mar \sin^2\theta}{\Sigma}, 
  \label{eq:metric_2}
\end{equation}
\begin{equation}
   g_{\varphi\varphi}
     \equiv 
     \frac{A \sin^2\theta}{\Sigma} , 
     \qquad
   g_{rr}
     \equiv 
     \frac{\Sigma}{\Delta} , 
     \qquad
   g_{\theta\theta}
     \equiv 
     \Sigma ;
  \label{eq:metric_3}
\end{equation}
$\Delta \equiv r^2-2Mr+a^2$,
$\Sigma\equiv r^2 +a^2\cos^2\theta$,
$A \equiv (r^2+a^2)^2-\Delta a^2\sin^2\theta$.
At the horizon, we obtain $\Delta=0$, 
which gives the horizon radius, 
$r_{\rm H} \equiv M+\sqrt{M^2-a^2}$,
where $M$ corresponds to the gravitational radius,
$r_{\rm g} \equiv GM c^{-2}=M$.
The spin parameter $a$ becomes $a=M$ for a maximally rotating BH,
and becomes $a=0$ for a non-rotating BH. 
The spacetime dragging frequency is given by
$\omega(r,\theta)= -g_{t\varphi}/g_{\varphi\varphi}$,
which decreases outwards as $\omega \propto r^{-3}$
at $r \gg r_{\rm g}=M$.

We assume that the non-corotational potential $\Phi$
depends on $t$ and $\varphi$ only through
the form $\varphi-\Omega_{\rm F} t$, and put
\begin{equation}
  F_{\mu t}+\Omega_{\rm F} F_{\mu \varphi}
  = -\partial_\mu \Phi(r,\theta,\varphi-\Omega_{\rm F} t) ,
  \label{eq:def_Phi}
\end{equation}
where $\Omega_{\rm F}$ denotes the magnetic-field-line
rotational angular frequency.
We refer to such a solution as a \lq stationary' solution
in the present paper.

The Gauss's law gives the Poisson equation
that describes $\Phi$ in a three dimensional magnetosphere
\cite{hiro06},
\begin{equation}
  -\frac{1}{\sqrt{-g}}
   \partial_\mu 
      \left( \frac{\sqrt{-g}}{\rho_{\rm w}^2}
             g^{\mu\nu} g_{\varphi\varphi}
             \partial_\nu \Phi
      \right)
  = 4\pi(\rho-\rho_{{\rm GJ}}),
  \label{eq:pois}
\end{equation}
where 
$\rho_{\rm w}^2 \equiv g_{t\varphi}^2 - g_{tt} g_{\varphi\varphi}
 = \Delta \sin^2\theta$,
and the general-relativistic Goldreich-Julian (GJ) charge density
is defined as
\cite{hiro06}
\begin{equation}
  \rho_{\rm GJ} \equiv 
      \frac{1}{4\pi\sqrt{-g}}
      \partial_\mu \left[ \frac{\sqrt{-g}}{\rho_{\rm w}^2}
                         g^{\mu\nu} g_{\varphi\varphi}
                         (\Omega_{\rm F}-\omega) F_{\varphi\nu}
                 \right].
  \label{eq:def_GJ}
\end{equation}
Far away from the horizon, $r \gg M$, 
equation~(\ref{eq:def_GJ}) reduces to the
ordinary, special-relativistic expression
of the GJ charge density~\cite{GJ69,mestel71},
\begin{equation}
  \rho_{\rm GJ} 
  \equiv -\frac{\mbox{\boldmath$\Omega$}\cdot\mbox{\boldmath$B$}}
               {2\pi c}
         +\frac{(\mbox{\boldmath$\Omega$}\times\mbox{\boldmath$r$})\cdot
                (\nabla\times\mbox{\boldmath$B$})}
               {4\pi c}.
  \label{eq:def_rhoGJ_1}
\end{equation}
Therefore, the corrections due to magnetospheric currents,
which are expressed by the second term of eq.~(\ref{eq:def_rhoGJ_1}),
are included in equation~(\ref{eq:def_GJ}).

If the real charge density $\rho$ deviates from the
rotationally induced Goldreich-Julian charge density,
$\rho_{\rm GJ}$, in some region,
equation~(\ref{eq:pois}) shows that
$\Phi$ changes as a function of position.
Thus, an acceleration electric field, 
$E_\parallel= -\partial \Phi / \partial s$,
arises along the magnetic field line,
where $s$ denotes the distance along the magnetic field line.
A gap is defined as the spatial region in which $E_\parallel$
is non-vanishing.
At the null charge surface,
$\rho_{{\rm GJ}}$ changes sign by definition.
Thus, a vacuum gap, 
in which $\vert\rho\vert \ll \vert \rho_{{\rm GJ}} \vert$,
appears around the null-charge surface,
because $\partial E_\parallel / \partial s$
should have opposite signs at the inner and outer boundaries
\cite{cheng86a,chiang92,romani96,cheng00}. 
As an extension of the vacuum gap, 
a non-vacuum gap,
in which $\vert\rho\vert$ becomes a good fraction of 
$\vert \rho_{{\rm GJ}} \vert$,
also appears around the null-charge surface
(\S~2.3.2 of HP~16),
unless the injected current across either the inner or the outer
boundary becomes a substantial fraction of the GJ value.

In previous series of our papers, e.g., \cite{hiro16b},
we have assumed $\Delta \ll M^2$ in Equation~(\ref{eq:pois}), 
expanding the left-hand side in the series of $\Delta/M^2$
and pick up only the leading orders.
However, in the present report, 
we discard this approximation, 
and consider all the terms that arise
at $\Delta \sim M^2$ or $\Delta \gg M^2$.

It should be noted that
$\rho_{\rm GJ}$ vanishes, and hence the null surface appears
near the place where $\Omega_{\rm F}$ coincides with 
the space-time dragging angular frequency, $\omega$ \cite{bes92}.
The deviation of the null surface
from this $\omega(r,\theta)=\Omega_{\rm F}$ surface is,
indeed, small, as figure~1 of \cite{hiro98} indicates.
Since $\omega$ can match $\Omega_{\rm F}$ only near the horizon,
the null surface, and hence the gap generally appears 
within one or two gravitational radii above the horizon,
irrespective of the BH mass. 

\section{Results}
\label{sec:5}
\begin{figure}
\includegraphics[width=10.0 cm]{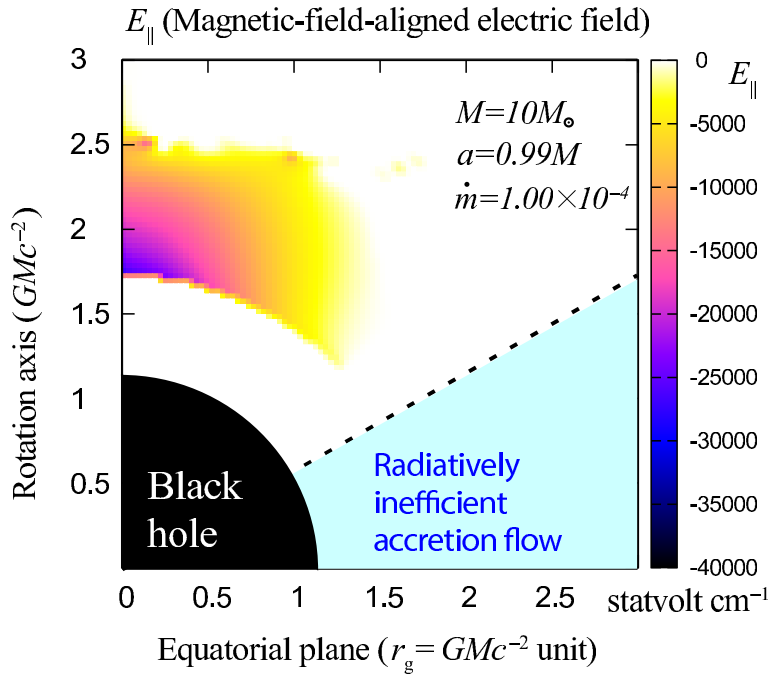}
\caption{
Magnetic-field-aligned electric field, $E_\parallel$, 
on the meridional plane. 
The filled black circle on the bottom left corner shows a black hole 
rotating along the ordinate. 
The mass and the spin parameter of the black hole are $M=10 M_\odot$ 
and $a=0.99 r_{\rm g}$. 
Both axes are normalized by the gravitational radius, 
$r_{\rm g}=GM c^{-2}$. 
The lepton accelerator appears only in the polar funnel, $\theta<60^\circ$. 
The dimensionless accretion rate is $\dot{m}=10^{-4}$. 
Magnetic field is assumed to be radial on the meridional plane, 
and be rotating with angular frequency $\Omega_{\rm F}=0.5 \omega_{\rm H}$, 
where $\omega_{\rm H}$ denotes the black hole’s spin angular frequency. 
The null-charge surface is located at radial coordinate, 
$r=1.73 r_{\rm g}$, whose $\theta$ dependence is weak.
}
\label{fig:2}
\end{figure}

We apply the method to a stellar-mass black hole with mass 
$M=10 M_\odot$. 
To consider an efficient emission, 
we consider an extremely rotating black hole, $a=0.99 r_{\rm g}$, 
because the accelerator luminosity rapidly increases 
as $a \rightarrow r_{\rm g}$ (10).
Owing to the frame-dragging effects, the Goldreich-Julian charge 
density decreases outwards around a rotating black hole. 
As a result, a negative $E_\parallel$ arises near the null-charge surface, 
which is located very close to the event horizon (fig.~\ref{fig:2}). 

\begin{figure}
\includegraphics[width=10.0cm]{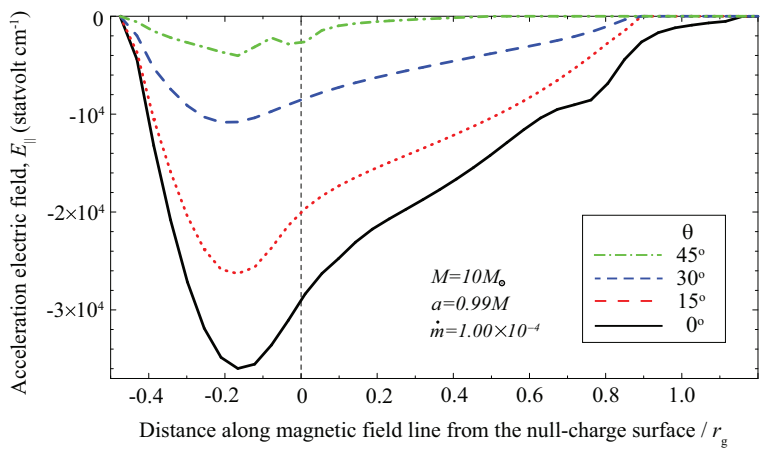}
\caption{
Distribution of the magnetic-field-aligned electric field, 
$E_\parallel$, that is presented in figure~1,
at four discrete colatitudes as labeled in the box,
where $\theta=0^\circ$ corresponds to the rotation axis.
The abscissa denotes the distance along the magnetic field
from the null-charge surface,
where the general relativistic Goldreich-Julian charge density
vanishes due to the spacetime dragging around a rotating black hole.
Black hole's mass ($M=10 M_\odot$),
spin ($a=0.99r_{\rm g}$), 
and the accretion rate ($\dot{m}=1.00 \times 10^{-4}$)
are common with figure~1.
The vertical dashed line shows the position of the null-charge surface
along $\theta=0^\circ$;
however, its position little depends on $\theta$
because we assume $\Omega_{\rm F}=0.5 \omega_{\rm H}$
(see the main text).
}
\label{fig:3}
\end{figure}

In fig.~\ref{fig:3}, we also plot $E_\parallel(r,\theta)$ 
at four discrete colatitudes,
$\theta=0^\circ$, $15^\circ$, $30^\circ$, and $45^\circ$.
It follows that $E_\parallel$ peaks slightly inside the
null surface (vertical dashed line),
and that it maximizes at $\theta=0^\circ$
(i.e., along the rotation axis).
The reason why $E_\parallel$ maximizes along the rotation axis
is that magnetic fluxes concentrate towards the
rotation axis
as the black hole spin approaches its maximum value
(i.e., as $a \rightarrow r_{\rm g}$)
\cite{komissa07,tchekhov10}.
Therefore, to consider the greatest gamma-ray flux,
we focus on the emission along the rotation axis,
$\theta=0^\circ$.
The acceleration electric field, $E_\parallel$,
decreases slowly outside the null surface
in the same way as pulsar outer gaps \cite{hiro99}.
This is because the two-dimensional screening effect of $E_\parallel$
works when the gap longitudinal (i.e., radial) width
becomes non-negligible compared to its trans-field 
(i.e., meridional) thickness. 

\begin{figure}
 \includegraphics[width=10.0 cm]{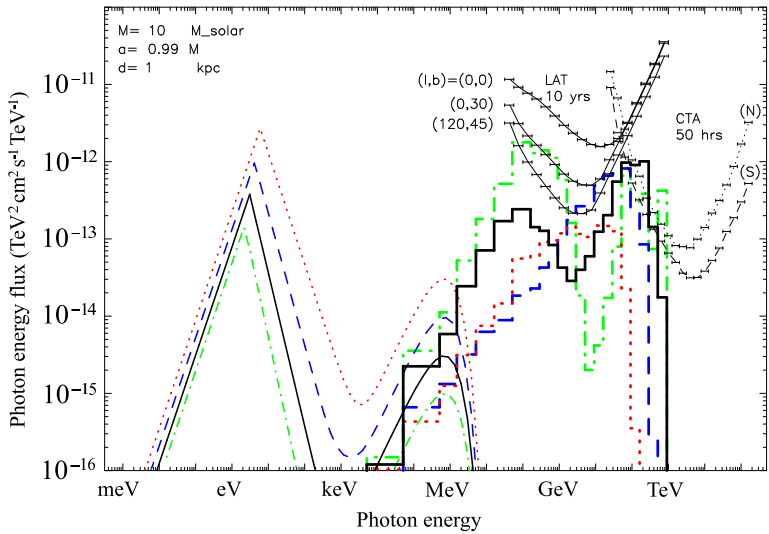}
\caption{
Spectrum of a black-hole lepton accelerator. 
The black hole mass and spin are common with figure~\ref{fig:1}. 
The red dotted, blue dashed, black solid, and green dash-dotted curves 
correspond to the dimensionless accretion rate of 
$10^{-3.50}$, $10^{-3.75}$, $10^{-4}$, and $10^{-4.25}$, respectively. 
The distance is assumed to be 1~kpc. 
The thin curves on the left denote the input spectra 
of the advection-dominated accretion flow, a kind of RIAF. 
Such soft photons illuminate the accelerator 
in the polar funnel. 
The thick lines denote the spectra of the gamma-rays emitted 
from the accelerator. 
The $0.1-10$~GeV photons are emitted via the curvature process, 
while those in $0.01-1$~TeV are via the inverse-Compton (IC) process. 
The detection limits of the Large Area Telescope (LAT) 
aboard the Fermi space observatory after ten-year observation  
are indicated by the thin solid curves. 
Also, the detection limits of the Cherenkov Telescope Array (CTA) 
after a 50-hour observation 
are shown by the thin dashed and 
dotted curves; 
(N) denotes the detection limit of the CTA in the northern-hemisphere, 
while (S) denotes those in the southern hemisphere.
}
\label{fig:4}
\end{figure}

The created $e^\pm$’s are accelerated by the $E_\parallel$ 
in opposite directions, emitting copious gamma-rays 
via the curvature process in $0.01-3$~GeV and 
via the IC process in $0.01-1$~TeV (fig.~\ref{fig:4}). 
The characteristic photon energy in the curvature process is given by
$h \nu_{\rm c}= (3/2) \hbar c \gamma^3 / \rho_{\rm c}$,
where $h$ denotes the Planck constant and
$\hbar \equiv h/2\pi$.
At each place in the gap, electrons have Lorentz factors typically
in the range $10^6 < \gamma < 3 \times 10^6$.
To evaluate the curvature radius $\rho_{\rm c}$, 
we assume that the horizon-threading magnetic field lines bend 
in the toroidal direction due to the frame dragging
and adopt $\rho_{\rm c}= r_{\rm g}$.
Since the pair production is sustained by the TeV photons
(emitted via the IC process), the gap electrodynamics is 
little affected by the actual value of $\rho_{\rm c}$,
which appears only in the curvature process.
Thus, we adopt this representative value, $\rho_{\rm c}= r_{\rm g}$.
The IC photon energy is limited by the electron kinetic energy
whose upper bound is about 1.5~TeV.
Thus, the IC photons have typical energies between 0.01~TeV and 1~TeV.

It also follows from figure~\ref{fig:4} that 
the gamma-ray luminosity increases with decreasing accretion rates. 
This is because the decreased RIAF soft photon field increases 
the pair-production mean-free path, the accelerator width 
along the magnetic field lines, and hence the electric potential drop. 
What is more, the emission becomes detectable with Fermi/LAT 
\footnote{LAT Performance 
 (https://www.slac.stanford.edu/exp/glast/groups/canda/lat\_Performance.htm)
}
and IACTs such as CTA
\footnote{CTA Performance 
(https://portal.cta-observatory.org/CTA\_Observatory/performance/SitePages/Hom.aspx)
}, 
if the distance is within 1~kpc, and if the dimensionless 
accretion rate resides in the narrow range,
$6 \times 10^{-5} < \dot{m} < 2 \times 10^{-4}$. 
The gamma-ray spectrum exhibits a turnover around TeV, 
because electron Lorentz factors are limited below $1.6$~TeV 
due to the curvature-radiation drag force. 
A caution should be made, however, on the assumption of
a stationary electron-positron pair cascade.
If the cascade takes place in a time-dependent manner
as suggested with Particle-in-Cell simulations \cite{levinson17},
the spectra might appear different from the present stationary analysis.
The present gap luminosity gives an estimate of the maximally possible
luminosity of a gap, 
whose electrodynamic structure may be variable in time. 
The dependence of the solutions on the BH spin will be discussed
in our subsequent paper \cite{hiro18}.

Let us analytically examine why the gap luminosity maximizes
when $\dot{m} \approx 10^{-4}$.
At radius $r$, the soft photon number density, $n_{\rm s}$, 
can be estimated as
\begin{eqnarray}
  n_{\rm s}
  &=& \frac{L_{\rm s}/c}{4\pi r^2 h\nu_{\rm s}}
  \nonumber\\
  &=& 1.3 \times 10^{20} M_1{}^{-2} 
      \left( \frac{10 r_{\rm g}}{r}
             \frac{d}{10 \mbox{kpc}} \right)^2
      \left( \frac{\mbox{eV}}{h\nu_{\rm s}}
             \frac{\nu_{\rm s} F_{\nu_{\rm s}}}
                  {\mbox{eV cm}^{-2} \mbox{ s}^{-1}} \right) 
      \, \mbox{cm}^{-3} ,
  \label{eq:ns} 
\end{eqnarray}
where $\nu_{\rm s} F_{\nu_{\rm s}}$ denotes the ADAF energy flux
whose value lies around
$10^{-12} \mbox{TeV cm}^{-2} \mbox{ s}^{-1}
 = \mbox{eV cm}^{-2} \mbox{ s}^{-1}$ 
at distance $d=10\mbox{ kpc}$;
the ADAF spectrum peaks at $h\nu_{\rm s} \approx \mbox{a few eV}$
(thin four curves on the left in fig.~4).
We evaluate the ADAF luminosity (in near-IR energies) with
$L_{\rm s} \approx 4\pi d^2 \nu_{\rm s} F_{\nu_{\rm s}}$, 
where $d= 10 \mbox{ kpc}$.
To compute $n_{\rm s}$, we assume that the photon density is
uniform within $r= 10 r_{\rm g}$,
a typical radius in which the equatorial ADAF is confined 
vertically by the magnetic pressure\cite{mckinney12}.

The electrons Lorentz factors are limited above $10^6$ \cite{hiro18}.
Thus, the Klein-Nishina cross section becomes 
$\sigma_{\rm IC} \approx 0.2 \sigma_{\rm T}$,
where $\sigma_{\rm T}$ denotes the Thomson cross section.
Thus, the mean-free path for the IC scatterings, 
$\lambda_{\rm IC}=1/(n_{\rm s} \sigma_{\rm IC})$,
becomes
\begin{equation}
  \frac{\lambda_{\rm IC}}{r_{\rm g}}
  \approx 4.0 \times 10^{-2} M_1
      \left( \frac{10 r_{\rm g}}{r}
             \frac{d}{10 \mbox{kpc}} 
      \right)^{-2}
      \left( \frac{\mbox{eV}}{h\nu_{\rm s}}
             \frac{\nu_{\rm s} F_{\nu_{\rm s}}}
                  {\mbox{eV cm}^{-2} \mbox{ s}^{-1}} 
      \right)^{-1} 
      \left( \frac{\sigma_{\rm IC}}{\sigma_{\rm T}}
      \right)^{-1} ,
  \label{eq:mfp_IC}
\end{equation}
The pair-production cross section becomes slightly below
$0.2 \sigma_{\rm T}$ for the collisions of TeV and eV photons
with moderate angles.
Thus, 
the sum of the IC and pair-production mean-free paths becomes
\begin{equation}
  \frac{\lambda_{\rm IC}+\lambda_{\rm pp}}{r_{\rm g}}
  \approx 2 \frac{\lambda_{\rm IC}}{r_{\rm g}}
  \approx 0.08 M_1
      \left( \frac{\mbox{eV}}{h\nu_{\rm s}}
             \frac{\nu_{\rm s} F_{\nu_{\rm s}}}
                  {\mbox{eV cm}^{-2} \mbox{ s}^{-1}} 
      \right)^{-1} ,
  \label{eq:l_IC_pp}
\end{equation}
Note that the Klein-Nishina and the pair-production cross sections 
are exactly computed in the numerical analysis,
taking account of the photon specific intensity 
and the particle distribution functions at each point.
Electrons are accelerated by $E_\parallel$ and attain 
the terminal Lorentz factor, $\gamma \sim 10^6$, 
after running the distance
\begin{equation}
 \lambda_{\rm acc} 
   = \frac{\gamma m_{\rm e} c^2}{e E_\parallel}
   = 1.7 \times 10^{5} 
     \left( \frac{\vert E_\parallel \vert}
                 {10^4 \mbox{ statvolt cm}^{-1}} 
     \right)^{-1}
     \frac{\gamma}{10^6} ,
  \label{eq:l_acc}
\end{equation}
which is less than $r_{\rm g}$.

We find that the gap becomes most luminous when 
\begin{equation}
 \lambda_{\rm IC}+\lambda_{\rm pp}+\lambda_{\rm acc}
   \approx \lambda_{\rm IC}+\lambda_{\rm pp}
   \approx r_{\rm g}=1.5 \times 10^6 M_1 \mbox{cm}.
  \label{eq:mfp_total}
\end{equation}
It follows from equation~(\ref{eq:l_IC_pp}) that
$\lambda_{\rm IC}+\lambda_{\rm pp} \approx 0.26 r_{\rm g}$ 
(or $\approx 0.40 r_{\rm g}$)
is realized when $\dot{m} \approx 10^{-4}$ 
(or $\approx 6 \times 10^{-5}$),
which gives $h\nu_{\rm s} \approx \mbox{eV}$ and
$ \nu_{\rm s} F_{\nu_{\rm s}} \approx 0.3$ 
(or $\approx 0.2$) $\mbox{eV cm}^{-2} \mbox{ s}^{-1}$.
It might appear that 
$\lambda_{\rm IC}+\lambda_{\rm pp} \approx r_{\rm g}$ 
holds if $\dot{m} \ll 10^{-4}$.
However, in this case,
the gap width rapidly increases to diverge;
that is, there exist no stationary solutions, 
(fig.~8 of \cite{hiro16a}).
Because of the simplification adopted around 
equations~(\ref{eq:ns})--(\ref{eq:mfp_total}),
we could not obtain 
$\lambda_{\rm IC}+\lambda_{\rm pp} \approx r_{\rm g}$ 
in this simplistic argument
in a consistent manner with the numerical results of
$E_\parallel$, $\gamma$, and the gamma-ray spectrum. 
Nevertheless, we can analytically conclude that
the gap longitudinal width becomes comparable to the horizon radius
and its luminosity maximizes when $\dot{m} \approx 10^{-4}$ 
for stellar-mass BHs.

To further analytically estimate 
$E_\parallel$, Lorentz factors, $\gamma$-ray energies, and so on
without invoking on the numerical results,
we have to perform the similar computations as described
in \S~2 of \cite{hiro13} for rotation-powered pulsars.
In this case, we would have to 
replace the curvature process with the IC process,
the neutron-star surface X-ray field with the ADAF IR field,
the neutron-star magnetic field with that created/supported by the ADAF,
and the light-cylinder radius with $r_{\rm g}$
(to compute the spatial gradient of $\rho_{\rm GJ}$).
For pulsars, the pair-production optical depth, $\tau_{\rm pp}$, 
is much less than unity for the out-going, curvature GeV photons,
which tail-on collide with the neutron-star surface X-rays.
However, for BHs, $\tau_{\rm pp} \sim 1$ holds for the IC TeV photons
whose collision angles are typically $0.5$--$1.0$~rad
with the ADAF-emitted near-IR photons.
It is noteworthy that the gap longitudinal width becomes 
approximately
$\lambda_{\rm IC}+\lambda_{\rm pp}+\lambda_{\rm acc}$,
because $\tau_{\rm pp} \sim 1$ holds for BH gaps.
It is, however, out of the scope of the present paper
to inquire the further details of this analytical method.

\section{Discussion}
\label{sec:6}
Let us compare the related gamma-ray emission scenarios. 
In the protostellar jet scenario
\cite{bosch10}, 
electrons and protons are accelerated 
at the termination shocks when the jets from massive protostars interact 
with the surrounding dense molecular clouds. 
Thus, the size of the emission region becomes comparable to 
the jet transverse thickness at the shock. 
In the hadronic cosmic ray scenario
\cite{ginz64,blandford87}, 
protons and helium nuclei 
are accelerated in the supernova shock fronts and propagate into 
dense molecular clouds, resulting in a single power-law photon spectrum 
in $0.001-100$~TeV through neutral pion decays. 
The size becomes comparable to the core of a dense molecular cloud. 
In the leptonic cosmic ray scenario
\cite{aharon97,swaluw01,hillas98}, 
electrons are accelerated 
at pulsar wind nebulae or shell-type supernova remnants, 
and radiate gamma-rays via IC process and radio/X-rays 
via synchrotron process.
Since the cosmic microwave background radiation provides 
the main soft photon field in the interstellar medium, 
the size may be comparable to the plerions, 
whose size increases with the pulsar age. 
In the black-hole lepton accelerator scenario
\cite{bes92,hiro98,neronov07,levinson11,globus14,brod15,hiro17,levinson17}, 
emission size does not exceed $10 r_{\rm g}$. 
Noting that the angular resolution of the CTA is 
about five times better than the current IACTs, 
we propose to discriminate the present black-hole lepton accelerator 
scenario from other scenarios by comparing the gamma-ray image and 
spectral properties. 
Namely, if a VHE source has a point-like morphology like 
HESS J1800-2400C in a gaseous cloud (section S1), 
and has two spectral peaks in $0.01-3$~GeV and 
$0.01-1$~TeV, 
but shows (synchrotron) power-law component 
in neither radio nor X-ray wavelengths, 
we consider that the present scenario accounts for its emission mechanism.

%
%
%




\end{document}